# A New Routing Protocol for All Optical Network

Kazi Sakib, Mosaddek Hossain Kamal and Ms. Upama Kabir
Network and Algorithm Research Lab
Department of Computer Science, University of Dhaka, Dhaka-1000,Bangladesh.
upama@udhaka.net, saykat_2000@yahoo.com

## ABSTRACT

*In this research paper, an efficient routing protocol for all optical network (AOL) is proposed. The technique uses wavelength division multiplexing (WDM). The proposed one is different from the conventional AOL protocol in transmission of data and control over optical fiber. A set of wavelengths is reserved to transfer control information, which is defined as control wavelengths. Control wavelengths are routed with packet routing scheme and the others are routed with wavelength routing scheme. In connection oriented network only the control packets are sent with the control wavelengths, data or messages are sent with other wavelengths. On the other hand, in datagram network, all the packets are sent with the control wavelengths. The allocation of wavelengths may be fully dynamic.*

## 1. INTRODUCTION

The fundamental purpose of the communication system was to exchange data between two stations, where the data meant only the text [1]. Nowadays, the data is expanded from text to audio-visual or multimedia. Moreover, multi-processing, parallel processing, multi-channels are very much common now. As a result, the need for speed and bandwidth is increasing. To fulfill the requirements, optical fiber communication is evolved. It will be seen in near future that, the whole backbone of the data communication system will only be built on optical fiber [2]. Though the optical network is the only solution that can fulfill the requirements, the existing networks cannot be thrown away. So, there should have some sort of compatibility between the optical networks and the existing electrical networks.

## 2. PRELUDE OF OPTICAL NETWORKS

In the Electro-Optical Network (EON), the optical data is sent to the router. The router converts the optical data to electrical domain and performs the routing decision and again converts it to optical domain, finally, the router sends it to its destination. At the destination the data is again converted to electrical domain [3]. An advantage of EON is that electronic routing can be performed. However, this advantage is also its potential downfall since the total throughput on the incoming links to the node must not exceed the electronic processing speed of the node. This has been called the electronic bottleneck [4].

The All-Optical Network (AON) is called the third generation of optical networks, where the data remains in optical domain from source to destination. No intermediate conversion is needed in AON [5]. The AON eliminates the electronic bottleneck; however since the electronic routing cannot be taken part, all-optical routing methods are required. One option is all-optical packet routing. There are severe technological problems with optical packet routing at this time. First of all, only the most basic of logic functions can be implemented. Second is the lack of the optical buffers [5]. The Broadcast-and-Select method can be used in AON. To have efficient throughput with this method a moderate number of wavelengths is required otherwise contention will be occurred. Moreover, the method has inherent power splitting loss, so it is not scalable [6]. Instead of all-optical packet routing, wavelength routing can be used. In a wavelength routing all-optical network (λ-routing AON), the path a signal takes is solely a function of the state of the devices, the wavelength of the signal and the location of the transmitter [7]. But in wavelength routing the number of wavelength is linearly proportional to the number of nodes in general [8]. This is why; in large networks sometimes another means of wavelength reusability is needed.

## 3. PROPOSED TECHNIQUE

It is mentioned that, the need for large amount of bandwidth gradually makes the communication system dependent on the optical fiber, but without a solid communication mechanism the whole potential of the optical fiber cannot be gained. Current optical routing protocols are suitable to some specialized cases. Many others are implemented in the Laboratory. They have a mere chance to come out at broad daylight. This research work tries to give a flavor of a new generalized approach for optical routing.

The problems of the existing techniques in both EON and AON are clearly portrayed before. Considering those problems and based on the criteria of ideal techniques, a new proposal is depicted. The proposed technique is textured with a novel network model first. Then it describes exclusively the circuit connection establishment procedure. To do so, modified all-optical routing node architecture is given. At the end, a set of network protocols is described for connection oriented and connectionless networks.

### 3.1 Network Model

Here the topology of an optical network is modeled as an undirected graph $G = (V, E)$ where each node in $V=\{A, B, C, D, E\}$ represents a router and each edge in E represents two fiber links, one in each direction, which is shown in the figure 1. Each node contains an injection buffer and a delivery buffer. Initially each message is stored in the injection buffer of its source. Once a message reaches its destination, it is stored in the destination's delivery buffer. During the routing, as the data will be routed with wavelength routing, storage is not needed actually for data. If anywhere the system fails to allocate any wavelength for a particular message, there should not be established any path in connection oriented method. Beyond this, the system assures that, if in connection oriented or in connection less method any data or message fails to get any wavelength, they will be discarded, which will be cost effective.

Each routing node contains collection of paths of the graph G, which are used as routing information. The routing node may use any routing strategies with WDM (Wavelength Division Multiplexing) [9]. But it must be remembered that, each connection between a pair of nodes in the network is assigned a path through the network. Moreover, wavelength on each link of that path is also assigned, such a way that, connections whose paths share a common link in the network are assigned different wavelengths.

The optical routing nodes are capable of routing each wavelength on an incoming link to any outgoing link. However, the same wavelength on two incoming links cannot be routed simultaneously onto a single outgoing link. That is, if there are W wavelengths on each link, the routing node may be viewed as consisting of W independent switches. Each node may have M inputs and outputs, that is have M incoming and outgoing links. In addition to routing and switching signals, the optical node also serves as source and sink of traffic in the network.

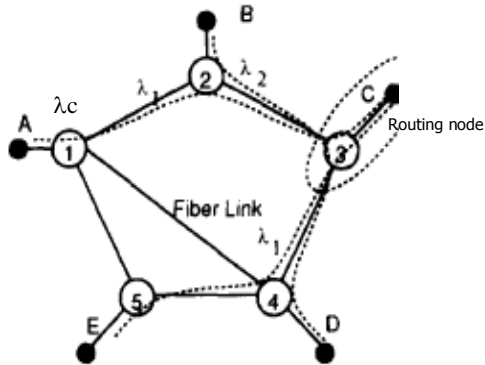
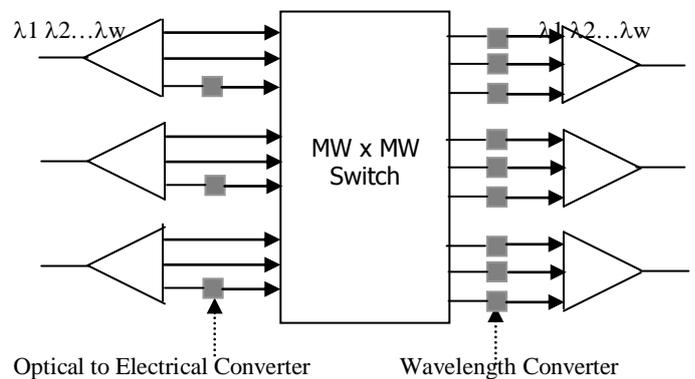

*Figure 1:* A WDM network consisting of routing node by point-to-point fiber optic link.

*Figure 2:* Proposed all-optical routing node *interconnected*

In this network model, connection requests and terminations may arrive randomly. Each connection must be assigned a specific path in the network and a specific wavelength for each links in the path provided that the wavelength conversion is allowed. Moreover the wavelengths and paths must be assigned such a way that two paths that share an edge should not be assigned the same wavelength [10].

**3.2 The Process of Connection Establishment**

The actual process of setting up routers and routes as well as wavelength assignment in optical networks is done using an electronic backbone control network [11]. In fact, the major applications for such networks require connections that last for relatively long periods once set up; thus the initial overhead is acceptable as long as sustained throughput at high data rates is subsequently available.

Interpreting the address header of messages arriving at optical switches is a serious problem, since their switching time is still slow compared to the transmission speed in optical fibers. In AT&T and elsewhere employs a low bit-rate header, which is read on the fly [12].

In the new technique, it is proposed that for requests and connection establishment, a subset λc, called control wavelengths, will be used and λc ⊆ B, where B is the set of wavelengths. That is for control, only a subset of wavelengths is used. Any node in the network wants to transmit data to another node, requests for a connection to its router with (λc) wavelength. With other information, the control packet or the request contains the wavelength information, by which the data will be sent. On the other hand, whenever the switch gets any message in wavelength (λc), it simply converts it in electrical domain and queued it up in its request queue. The router or switch processes the request through establishing a path to the next hop, and assigning an unused wavelength for that connection. Then the router informs the source to send the messages (upon assigned priority) or after a pre-specified delay the source starts to send data if it is a non-priority network. In the mean time the router gets prepared to catch that data with that pre-specified wavelength. The router sends the unconverted data to the next hop, along the proper outgoing path. The next hop performs the same tasks and sends the messages to its next hop and thus to the destination. If the router cannot specify a suitable wavelength for the messages for its next outgoing link, it simply discards the messages. On the other hand, for datagram communication all the messages should be sent with the control wavelengths.

Number of the control wavelengths may be varied. That is, the optical routing nodes may dynamically increase or decrease the number of control wavelengths with the increasing or decreasing of requests.

### 3.3 Proposed All-Optical Routing Node

In this section a new routing node is proposed for the newer technique. The specialty of the node is that, it only converts the messages coming with control wavelength (λc) into the electrical domain, and the messages coming with other wavelengths, just let them pass to their next hop without converting them to the electrical domain. Figure 2 depicts the proposed routing node for all-optical network. It is based on a large optical switch, which takes a channel and switches it to any other channel (on any fiber). Before being multiplexed into the fiber, each channel is converted to the appropriate wavelength by fixed wavelength converters.

### 3.4 Protocol

## 3.4.1 Protocol for Connection Oriented AON

It is stated above that, whenever a node wants to transmit anything, it first have to request the routing node with control wavelengths. After receiving the confirmation from the routing node or after a pre-specified time, the source node will send its messages with specified wavelength. The protocol for the source side is stated below.

Source:
- Send all requests to the routing node with λc
- Wait for reply from the routing node
- After having the reply send the messages with selected wavelength
- Wait for final acknowledgement from the destination
- Delete the message from the injection buffer

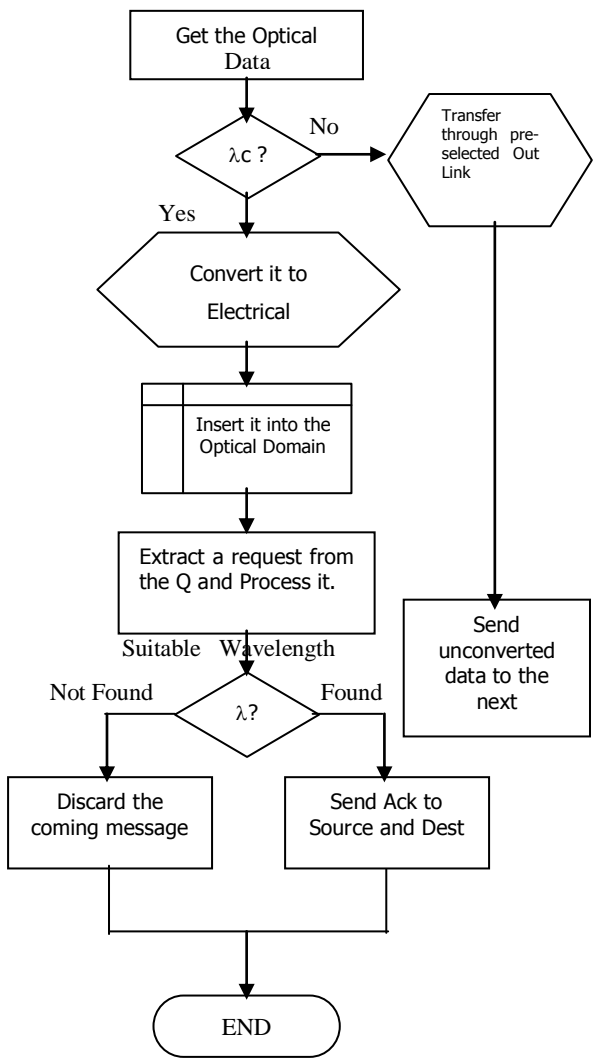

*Figure 3: Flow chart for protocol of routing node.*

The routing node or switch is always waiting for requests from the attached nodes. It performs the routing decision and allocates a wavelength for its outgoing link for each request. It passes the data coming with other wavelengths than the control wavelengths directly to the outgoing link without converting the data in electrical domain. It does not store those data anymore.

Routing Node:
- Wait for data/messages/requests from the source
- If it is with $\lambda c$
    - Convert it in electrical domain
    - Put it in the buffer/queue
    - Extract a request/packet from the buffer/queue and process it
    - Readjust the incoming link receiver (and send a reply to its source)
    - Assign a wavelength to the outgoing link and send the request to its next hop
- Otherwise select proper outgoing link and send the unconverted data to the next hop, with wavelength conversion if necessary.
    - If suitable wavelength cannot be assigned discard the message.

A flow chart of the routing node protocol is seen in the figure 3.

Destination:

Store the data in the delivery buffer

Send an acknowledgement to the Source.

### 3.4.2 Protocol for Connectionless AON

In datagram network all the messages are send with control wavelengths. As the network is conflict free, all the messages must arrive in their destination.

Source:

- Send all the messages with $\lambda c$
- Delete the messages from the injection buffer

Routing Node:
- Wait for data/messages/requests from the source
    - If it is with $\lambda c$
    - Convert it in electrical domain

- Put it in the buffer/queue
- Extract a request/packet from the buffer/queue and process it
- Read the header and send it to its next hop

Destination:

- Store the data in the delivery buffer

**4. Effects of the Newer Technique**

In this technique the routing node assigns suitable wavelength for each single link, for every data or message, so the system is fully contention free. Moreover no extra connection establishment electronic backbone control network is needed, thus it is simple. The given routing node or switch can easily create the connection with the control wavelengths, which are shown. The switch is dynamic. It can assign or discard any wavelength to control wavelength set, which helps to reduce the rate of message discarding. The most flexibility of the network is, it can use the packet routing as well as the wavelength routing. Actually, the network uses a blend of packet and wavelength routing technique, which will give it a better outcome.

Though the proposed protocol seems simple, but it needs a new routing node or switch. The structure of the switch is little bit complex. To make the switch dynamic, it is needed that, each of the incoming link of the switch must have the electronic converter. At the same time the routing node must have enough storage to store the control packets. Control packets for connection-oriented network, are not a big deal, but the datagram packets are generally huge in both numbers and amounts. Moreover, the routing node has full wavelength conversion capability to make the network most flexible, which increases the complexity as well as the cost. Despite of those, the switch can be made compatible to other techniques and protocols. As the router supports packet routing to wavelength routing, the network can handle all the first to third generation optical transmissions.

It is stated that, the EON can perform the electronic routing e.g., virtual circuit or packet routing, but it makes the potential downfall for the electronic bottlenecks [4]. The new technique can provide the packet routing, with minimizing the electronic bottlenecks at the least. The new proposed model doesn't contain any intermediate message buffers, and it assures that, it will discard least number of messages. Unlike the broadcast-and-select network, here flooding technique is never used. Instead of all-optical packet routing, the recent all-optical networks use the wavelength routing technique. Here in the new technique, wavelength routing also used and the bottlenecks of the wavelength routings are still alive here.

## 5. SIMULATIONS AND PERFORMANCE EVALUATION

At the end of the research work a simulation has been performed. The simulation program compares an existing method (for example: SONET/SDH) and the new proposed technique. The simulation program is run under some fixed parameters. The request size is taken 1000. It is assumed that, a single request contains 100 flits or packets. The propagation delay is taken 1 $\mu$ second and the switch processing time is assumed 2 $\mu$ second.

The required time, of a single request with its all messages to pass a single switch, is taken into account. The time is taken for a fixed number of requests. For some specified wavelengths and control wavelengths, the degree of parallelism is varied. Thus for various number of wavelengths, as well as control wavelengths, the result is taken. Then the result is compared.

The achieved results from the two simulated protocols are given in tabular and graphical form. Each case is evaluated by considering the number of degree of parallelism and corresponding processing time.

**Case-1:** Table 1, shows the required time for 100 requests or 10000 packets to pass a single switch of the existing protocol and the proposed protocol. The values are taken in $\mu$ seconds and are found under the following conditions – number of wavelengths is 4 and the control wavelength number is 1. The graphical representation of the table is given in the graph 1 which shows that the proposed technique acquires much lower time than the existing one.

**Number of Wavelengths – 4, Control Wavelengths – 1**

| Degree of Parallelism | Existing Protocol (required time to pass the switch in $\mu$ seconds) | Proposed Protocol (required time to pass the switch in $\mu$ seconds) |
|---|---|---|
| 1 | 385147 | 10701 |
| 2 | 99412 | 2675 |
| 3 | 59734 | 1337 |
| 4 | 40482 | 668 |

**Table 1:** Performance Evaluation Of Existing Protocol and Proposed Protocol

**Case-2:** In table 2 the results are found, when the wavelength number is increased to 16, the number of control wavelengths is 4 now. The requests number is taken 100 again. The graphical representation of the table is shown in graph 2.

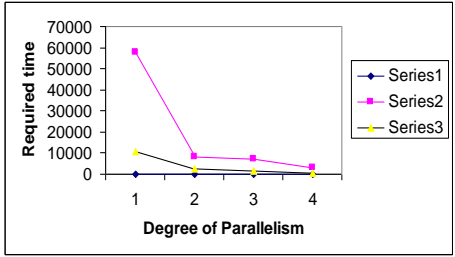
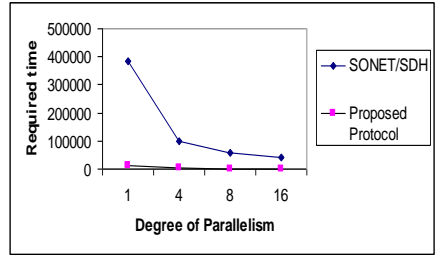
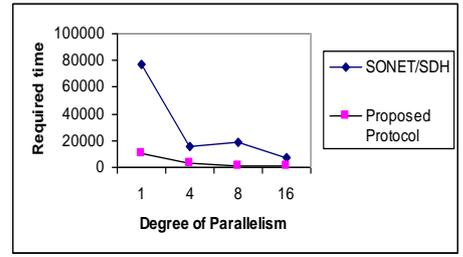

**Graph 1:** graphical representation of table 1.    **Graph 2:** graphical representation of table 2    **Graph 3:** graphical representation of table 3

Number of Wavelengths – 16, Cntrl Wavelengths – 4

| Degree of Parallelism | Existing Protocol (Required time to pass the switch in μ seconds) | Proposed Protocol (required time to pass the switch in μ seconds) |
|---|---|---|
| 1 | 76599 | 10701 |
| 4 | 15382 | 2675 |
| 8 | 18234 | 1337 |
| 16 | 7136 | 668 |

**Table 2:** Performance Evaluation Of Existing Protocol and Proposed Protocol

Number of Wavelengths – 64, Cntrl Wavelengths – 16

| Degree of Parallelism | Existing Protocol (required time to pass the switch in μ seconds) | Proposed Protocol (required time to pass the switch in μ seconds) |
|---|---|---|
| 1 | 57843 | 10701 |
| 4 | 8322 | 2675 |
| 8 | 7199 | 1337 |
| 16 | 2997 | 668 |

**Table 3:** Performance Evaluation Of Existing Protocol and Proposed Protocol

Graph 2 and table 2 shows that the result for existing protocol is improving but they are far behind from the proposed protocol.

**Case-3:** Table 3 is composed under following conditions – the wavelength number is now 64, and the control wavelength is now on 16. Like previous cases, the degree of parallelism is varied and the load is fixed, i.e., 100 requests or 10000 packets. The graphical representation of the table is given in graph 3.

At the end of the performance evaluation it can be said that the newer protocol is far ahead from the existing technique. Though the existing protocol's performance dramatically increases if the degree of parallelism is increased, it is behind well enough than the newer one. In all the three cases, the tables and graphs, given above, can provide almost clear view of the difference of performances between the previous technique and the newer.

## 6. CONCLUSION

During the research work, different classes of wavelength division multiplexing (WDM) all-optical networks are studied. The electro-optical networks are also taken into account. For various networks various networking techniques or protocols are found. Among them the widely used protocols are taken. For example, the SONET/SDH of the EON, the "Trial and Failure protocol" of the AON. Those protocols are analyzed and simulated. The result shows the limitations of the existing techniques. In this work a new protocol for wavelength division multiplexing (WDM) all-optical network is proposed and also its effects on the network is shown. It is believed that the main contribution of this research has been as much in opening the door of a new arena. During the research period, at first the criteria of an ideal optical networking technique or protocol are fixed. Depends on those criteria, the new generalized protocol is proposed, which is simple but efficient. Moreover the protocol is fully conflict-free.